\documentclass[10pt, conference]{IEEEtran}
\IEEEoverridecommandlockouts
\usepackage{cite}
\usepackage{amsmath,amssymb,amsfonts}
\usepackage{algorithmic}
\usepackage{graphicx}
\usepackage{textcomp}
\usepackage{xcolor}

\def\BibTeX{{\rm B\kern-.05em{\sc i\kern-.025em b}\kern-.08em
    T\kern-.1667em\lower.7ex\hbox{E}\kern-.125emX}}
\begin{document}

\title{Profit Maximization In Arbitrage Loops}

\author{\IEEEauthorblockN{Yu Zhang}
\IEEEauthorblockA{ \textit{University of Zurich, Switzerland}\\ 
Email: zhangyu@ifi.uzh.ch}
\and
\IEEEauthorblockN{Zichen Li}
\IEEEauthorblockA{ \textit{Chinese University of Hong Kong, China}\\ 
Email: zichenli@link.cuhk.edu.cn}
\and
\IEEEauthorblockN{Tao Yan}
\IEEEauthorblockA{\textit{University of Zurich, Switzerland}\\ 
Email: yan@ifi.uzh.ch}
\and
\IEEEauthorblockN{Qianyu Liu}
\IEEEauthorblockA{ \textit{University of Zurich, Switzerland}\\ 
Email: qianyu@ifi.uzh.ch}
\and
\IEEEauthorblockN{Nicolo Vallarano}
\IEEEauthorblockA{ \textit{University of Zurich, Switzerland}\\ 
Email: nicolo.vallarano@uzh.ch}
\and
\IEEEauthorblockN{Claudio J. Tessone}
\IEEEauthorblockA{ \textit{University of Zurich, Switzerland}\\ 
Email: tessone@ifi.uzh.ch}
}

\maketitle

\begin{abstract}
Cyclic arbitrage chances exist abundantly among decentralized exchanges (DEXs), like Uniswap V2. For an arbitrage cycle (loop), researchers or practitioners usually choose a specific token, such as Ether (the cryptocurrency from Ethereum) as input, and optimize their input amount to get the net maximal amount of the specific token as arbitrage profit without considering the tokens' market price from the centralized markets (CEXs). 
By considering the tokens' prices from CEXs in this paper, the new arbitrage profit will be quantified as the product of the net number of a specific token we got from the arbitrage loop and its corresponding price in CEXs. The new arbitrage profit will be called monetized arbitrage profit in this paper.
Based on this concept, we put forward three different strategies to maximize the monetized arbitrage profit for each arbitrage loop.
The first strategy is called the $MaxPrice$ strategy. Under this strategy, arbitrageurs start arbitrage only from the token with the highest CEX price.
The second strategy is called the $MaxMax$ strategy. Under this strategy, we calculate the monetized arbitrage profit for each token as input in turn in the arbitrage loop. Then, we pick up the most maximal monetized arbitrage profit among them as the monetized arbitrage profit of the $MaxMax$ strategy. It is easy to prove that this strategy can bring more profit than the $MaxPrice$ strategy.
The third one is called the $Convex \: Optimization$ strategy. In this strategy, we mapped the $MaxMax$ strategy to a convex optimization problem and proved that the $Convex \: Optimization$ strategy could get more profit in theory than the $MaxMax$ strategy, which is proved again in a given example. 
We also proved that if no arbitrage profit exists according to the $MaxMax$ strategy, then the $Convex \: Optimization$ strategy can not detect any arbitrage profit, either. 
However, the empirical data analysis denotes that the profitability of the $Convex \: Optimization$ strategy is almost equal to that of the $MaxMax$ strategy, and the $MaxPrice$ strategy is not reliable in getting the maximal monetized arbitrage profit compared to the $MaxMax$ strategy.
\end{abstract}


\section{Introduction} 
Decentralized finance (DeFi) has emerged as a promising field within the cryptocurrency ecosystem, leveraging blockchain technology to create financial applications that are open, transparent, and accessible to anyone. One key part of DeFi is decentralized exchanges (DEXs), which facilitate peer-to-peer trading of tokens without the need for intermediaries \cite{Jani2022AnEmp}. Traders engage in transactions either amongst themselves or against a token liquidity pool, exchanging one token for another within the decentralized exchanges (DEXs). 

Among various DEXs, Uniswap is the largest DEX in terms of Total Value Locked (TVL) running on the Ethereum blockchain. Uniswap adopts the Automated Market Maker (AMM) mechanism\cite{10.1145/3570639}, which facilitates tokens' trading using a liquidity pool without an intermediary market maker. 
Anyone can participate in the transaction activity on the Uniswap platform without approval from a centralized authority. There are two main kinds of participants on this platform, liquidity providers and liquidity takers. Liquidity providers are those who build a new liquidity pool or inject liquidity into an already existing liquidity pool. Liquidity takers, also called traders in this paper, are those who apply the liquidity pool to exchange tokens.

Under the AMM model, if a pair of tokens form a liquidity pool, then this pair of tokens can be exchanged with each other and the relative price (exchange rate) of this pair of tokens equals the ratio of their reservation balance in the same pool. The Uniswap V2 uses the constant product market maker (CPMM) for cryptocurrency exchange, which leads to price slippage that limits the arbitrage profit that traders can get.

As of October 2023, there have been more than ten thousand tokens and one hundred thousand liquidity pools by checking the on-chain data. Similar to the foreign exchange market (Forex) where triangular arbitrage\cite{cui2020detecting} exploits exchange rate differences between different currency pairs, extensive research has shown that Uniswap provides abundant arbitrage opportunities due to the price discrepancies of tokens across various liquidity pools \cite{Jani2022AnEmp,yewa2023cyclic}. For example, \cite{yewa2023cyclic} found that the revenue of the most exploitable arbitrage opportunity is higher than 1 Ether for each block from May 2020 to April 2021, and almost three hundred thousand loop transactions within eleven months. \cite{zhou2021just} also used the Bellman-Ford-Moore algorithm to detect the arbitrage loops.

After an arbitrage loop is found, researchers usually only target a specific token, like Ether, to calculate the corresponding arbitrage profit. We call this strategy the traditional strategy in this paper. For example, \cite{yewa2023cyclic, zhou2021just} took this strategy to calculate the arbitrage profit starting and backing to Ether for all arbitrage loops.

By considering each token's price from centralized exchanges (CEXs), we introduced the concept of monetized arbitrage profit which is quantified as the product of the net number of a specific token we got from the arbitrage loop as profit and its corresponding price in CEXs. For example, for an arbitrage loop starting from token $A$ and back to $A$ again ($A \rightarrow B\cdots \rightarrow A$), if we get $\pi_A$ unit of token $A$ as profit at most by input token $A$ in the arbitrage loop and its price in CEXs is $P_A$, then the monetized arbitrage profit of starting from token $A$ is $\pi_A  P_A$. If we start arbitrage from different tokens, then the monetized arbitrage profits are also different (e.g. $\pi_B  P_B$).
So, after arbitrage loops were found, a question is whether they got maximal monetized arbitrage profits.

Based on this consideration, we put forward several different strategies to maximize the monetized arbitrage profit for each arbitrage loop.
Researchers or practitioners may also think that the maximal monetized arbitrage profit can be obtained if arbitrageurs always start arbitraging from the token with the highest CEX price in the arbitrage loop. We call this strategy the $MaxPrice$ strategy which is just like the traditional arbitrage strategies in the implementation process and is our first strategy. We will show that the $MaxPrice$ is not reliable in getting the maximal monetized arbitrage profit in an example and by empirical data from Uniswap V2.
The second strategy is called the $MaxMax$ strategy. Under this strategy, not only one specific token (like Ether or another token with the highest CEX price), but each token in the arbitrage loop will be taken as input in turn, and then we calculate their respective monetized arbitrage profit. At last, we pick up the most maximal monetized arbitrage profit among them as the monetized arbitrage profit of the $MaxMax$ strategy ($Max(\pi_A  P_A,\pi_B  P_B,\dots)$). It is easy to prove that this strategy can bring more profit than the traditional strategy and the $MaxPrice$ strategy because $Max(\pi_A  P_A,\pi_B  P_B,\dots)\geq \pi_A  P_A$, $Max(\pi_A  P_A,\pi_B  P_B,\dots)\geq \pi_B  P_B$, $\dots$.
The third one is called a $Convex \: Optimization$ strategy. In the later section, we will map the $MaxMax$ strategy to a convex optimization problem and prove that the $Convex \: Optimization$ strategy could get more profit in theory than the $MaxMax$ strategy, which will be shown again in a given example. 
We will also prove that if no arbitrage profit exists according to the $MaxMax$ strategy, then the $Convex \: Optimization$ strategy can not detect any arbitrage profit, either. 

The paper is organized as follows: Section 2 introduces related work about arbitrage on Uniswap; Section 3 introduces the AMM rules and the $MaxMax$ strategy to get more monetized arbitrage profit along the arbitrage loops. Section 4 shows how to map the $MaxMax$ strategy to the $Convex \: Optimization$ strategy and prove that the monetized profit we get by the $Convex \: Optimization$ strategy is larger than the $MaxMax$ strategy. Section 5 gives an arbitrage example and shows that the $MaxMax$ strategy can get more monetized profit than the traditional strategy, but
the $Convex \: Optimization$ strategy can get more monetized profit than the $MaxMax$ strategy. We will also show that the $MaxPrice$ strategy is not liable in calculating the maximal monetized arbitrage profit. In Section 6, we analyze some empirical loop arbitrage data on Uniswap V2 and compare the profitability of traditional strategies, the $MaxMax$ strategy to that of $Convex \: Optimization$ strategy, the traditional strategies, and the $MaxPrice$ strategy. Section 7 summarizes and discusses the relative advantages and disadvantages of the $MaxMax$ strategy and the $Convex \: Optimization$ strategy.

\section{Related Work} 

The arbitrage on a DEX is similar to the Forex in the aspect that any exchangeable pair of tokens has a unique exchange rate, but also different from the arbitrage in the Forex, like in the exchange rate determination process.

After the appearance of the DEXs, like Uniswap V2, though still in its early stage, there have been some scientific papers that focus on analyzing the arbitrage opportunities of DEXs. Wang et al.\cite{wang2022cyclic} analyze the potential cyclic arbitrage chances and explored arbitrage profit by traversing all triangles with Ether (ETH), the native token of Ethereum, included on Uniswap V2. Robert et al.\cite{mclaughlin2023large} used the trade event log to recognize historic arbitrage profit and applied Johnson’s cycle-detection algorithm to look for potential arbitrage chances. Zhou et al.\cite{zhou2021just} applied the Moore-Bellman-Ford algorithm to recognize arbitrage loops. Danos et al.\cite{danos2021global} took the arbitrage problem as a convexity problem and applied the optimization operation to find arbitrage paths from a theoretical perspective. Berg et al. \cite{berg2022empirical} applied the method in \cite{danos2021global} on Uniswap V2 to research the efficiency of DEX by recognizing profitable arbitrages.

In this paper, we do not focus on the arbitrage loop detection on DEXs, but on how to maximize the monetized arbitrage profit in an arbitrage loop, which makes our research very different from those that
usually only target a specific token without considering tokens' price in CEXs\cite{wang2022cyclic, zhou2021just}. \cite{danos2021global} introduce the convex optimization method to detect the arbitrage path and calculate traders' optimal utility. Using the convex optimization method to detect the arbitrage paths, like \cite{danos2021global}, can be problematic because the arbitrage paths we get are always very complicated and are hard to implement in the real Uniswap platform.
However, their work enlightens our research in using the $Convex \: Optimization$ strategy to calculate the monetized arbitrage profit for an arbitrage loop. Our work focuses on the strategies to maximize the monetized arbitrage profit for any arbitrage loops, but not an arbitrage detection problem, which makes our research very different from that in \cite{danos2021global}. There is still a lack of research systematically focusing on the maximization of monetized arbitrage profit in arbitrage loops on Uniswap.

\section{AMM Exchange Rules in DEXs and the $MaxMax$ Strategy}\label{traditional}

Firstly, we define the relative price between two tokens in the same liquidity pool. We use $p_{ij}$ to denote the price of token $i$ in the unit of token $j$ which is the ratio between the reserve of the token $i$ and $j$ in the same pool. So, $p_{ij}=(1-\lambda)\frac{r_j}{r_i}$, where $r_i$ and $r_j$ is the reserve of token $i$ and $j$ , and $\lambda$ is the imposed tax rate by the liquidity pool. Correspondingly, $p_{ji}$ is the price of token $j$ in the unit of token $i$ and $p_{ji}=(1-\lambda)\frac{r_i}{r_j}$.
A token loop is called an arbitrage loop when the product of all relative prices along the token loop is larger than 1. For example, in the case of three tokens in the same loop, when $p_{ij} \cdot p_{jk} \cdot p_{ki} >1$, or $log(p_{ij}) + log(p_{jk}) + log(p_{ki})>0$, this loop is an arbitrage loop, where $i$, $j$, $k$ are three different tokens. This condition is applicable in the case of more tokens in an arbitrage loop.

After we find an arbitrage loop, the next question is how much we should invest to maximize our profit. Now, we explain it as follows.

The constant AMM equation in any liquidity pool of Uniswap V2 is in the form of
\begin{displaymath}
        [x+(1-\lambda) \Delta x](y-\Delta y) = xy=k,
\end{displaymath}
where $x$ and $y$ are constant and denote the reserve of token $X$ and $Y$ before trading in the liquidity pool, respectively. $k$ equals the product of $x$ and $y$, $\lambda$ is the transaction tax rate. $\Delta x$ and $\Delta y$ denote the input number of token $X$ and the output number of token $Y$ in trading, which are variables that we will focus on. So, the above equation quantifies the number of token $Y$ we can get ($\Delta y$) by inputting a specific number of token $X$ ($\Delta x$) in trading. By simple derivation, we can get the function $\Delta y = F(\Delta x|\theta) $ between $\Delta y$ and $\Delta x$:
\begin{displaymath}
    F(\Delta x|\theta)= y-\frac{x\cdot y}{x+(1-\lambda)\cdot \Delta x}
\end{displaymath}
where $\theta = (x,y,\lambda)$ denotes a parameter tuple including $x$, $y$ and $\lambda$.
From the above equation, we know that $\Delta y = F(\Delta x|\theta) $ is a convex and monotone-increasing function with $\Delta x$. When an arbitrage path includes more tokens, the number of inputs of the starting token is still a convex and monotonically increasing function of the number of target tokens taken out.

Now,  we assume there exists a cyclic arbitrage opportunity by exchanging token $X$ to token $Y$, then from token $Y$ to token $Z$, and finally from token $Z$ to token $X$ again, namely $X\rightarrow Y \rightarrow Z \rightarrow X$. 

In the first liquidity pool between token $X$ and token $Y$, the function between $\Delta y$ and $\Delta x$ is:
\begin{equation}
    \Delta y = F(\Delta x|\theta_1)= y_1-\frac{x_1\cdot y_1}{x_1+(1-\lambda)\cdot \Delta x}
    \label{e1}
\end{equation}
where $\theta_1 = (x_1,y_1,\lambda)$ denotes the parameters in the first liquidity pool, $x_1$ and $y_1$ is corresponding reserve of token $X$ and $Y$ in the first liquidity pool, respectively.

In the second liquidity pool between token $Y$ and token $Z$, the function between $\Delta z$ and $\Delta y$ is:
\begin{equation}
    \Delta z= F(\Delta y|\theta_2)= z_2-\frac{y_2\cdot z_2}{y_2+(1-\lambda)\cdot \Delta y}
    \label{e2}
\end{equation}
where $\theta_2 = (y_2,z_2,\lambda)$ denotes the parameters in the second liquidity pool, $y_2$ and $z_2$ is corresponding reserve of token $Y$ and $Z$, respectively.

In the third liquidity pool between token $Z$ and token $X$, the function between $\Delta x$ and $\Delta z$ is:
\begin{equation}
    \Delta x= F(\Delta z|\theta_3)= x_3-\frac{z_3\cdot x_3}{x_3+(1-\lambda)\cdot \Delta z}
    \label{e3}
\end{equation}
where $\theta_3 = (z_3,x_3,\lambda)$ denotes the parameters in the third liquidity pool, $z_3$ and $x_3$ is corresponding reserve of token $Z$ and $X$, respectively.
To differentiate the input amount of token $X$ in equation \ref{e1} from the output amount of token $X$ in equation \ref{e3}, we use $\Delta x_{in}$  and $\Delta x_{out}$  to denote the input amount in equation \ref{e1} and output amount in equation \ref{e3}, respectively.
The problem of maximizing profit in this arbitrage loop is
\begin{equation}
    Max(\Delta x_{out} - \Delta x_{in}).
    \label{e4}
\end{equation}

By equation \ref{e1}, \ref{e2} and \ref{e3}, $\Delta x_{out}=F(F(F(\Delta x_{in}|\theta_1)|\theta_2)|\theta_3)$ and the fact that $F(\cdot)$ is convex and increases monotone, the maximization problem will be a straightforward one-variable maximization problem, namely, $Max[F(F(F(\Delta x_{in}|\theta_1)|\theta_2)|\theta_3)-\Delta x_{in}]$
Similarly, by equation \ref{e1}, \ref{e2} and \ref{e3}, the objective function to maximize from token $Y$ and $Z$ are $\Delta y_{out}=F(F(F(\Delta y_{in}|\theta_2)|\theta_3)|\theta_1)$ and $\Delta z_{out}=F(F(F(\Delta z_{in}|\theta_3)|\theta_1)|\theta_2)$, respectively. These definitions of $\theta_1$, $\theta_2$ and $\theta_3$ are the same as that in equations \ref{e1}, \ref{e2} and \ref{e3}. The only difference among $\Delta x_{out}$, $\Delta y_{out}$ and $\Delta z_{out}$ is the different orders of $\theta_1$, $\theta_2$ and $\theta_3$ in the functions. 

If we input a small amount of token $X$ ($\Delta x_{in}^,$) based on equation \ref{e4}, then the reserves of tokens in the liquidity pool along the loop will change, which means that these parameters in functions $\Delta y_{out}$ and $\Delta z_{out}$ will change. The question is how the change of these parameters affects the functions of $\Delta y_{out}$ and $\Delta z_{out}$. After we input a small amount of token $X$ and get some arbitrage profit in the arbitrage loop $X\rightarrow Y \rightarrow Z \rightarrow X$, the total potential arbitrage profit in the same loop will decrease, which means that the arbitrage profit in the loop $Y\rightarrow Z \rightarrow X \rightarrow Y$ by inputting token $Y$ and $Z\rightarrow X\rightarrow Y \rightarrow Z$ by inputting token $Z$ will decrease, too. This denotes that the input in token $X$ will lead to the smaller $\Delta y_{out}$ and $\Delta z_{out}$ for each value of $y_{in}$ and $z_{in}$, respectively. 

When we increase the input amount of token $X$ until the margin input of $\Delta x_{in}$ equals the margin output of $\Delta x_{out}$ in the amount equals, namely $\frac{d\Delta x_{out}}{d\Delta x_{in}}=1$, the profit is maximal.

\begin{figure}[!ht]
    \centering
    \includegraphics[width=0.4\textwidth]{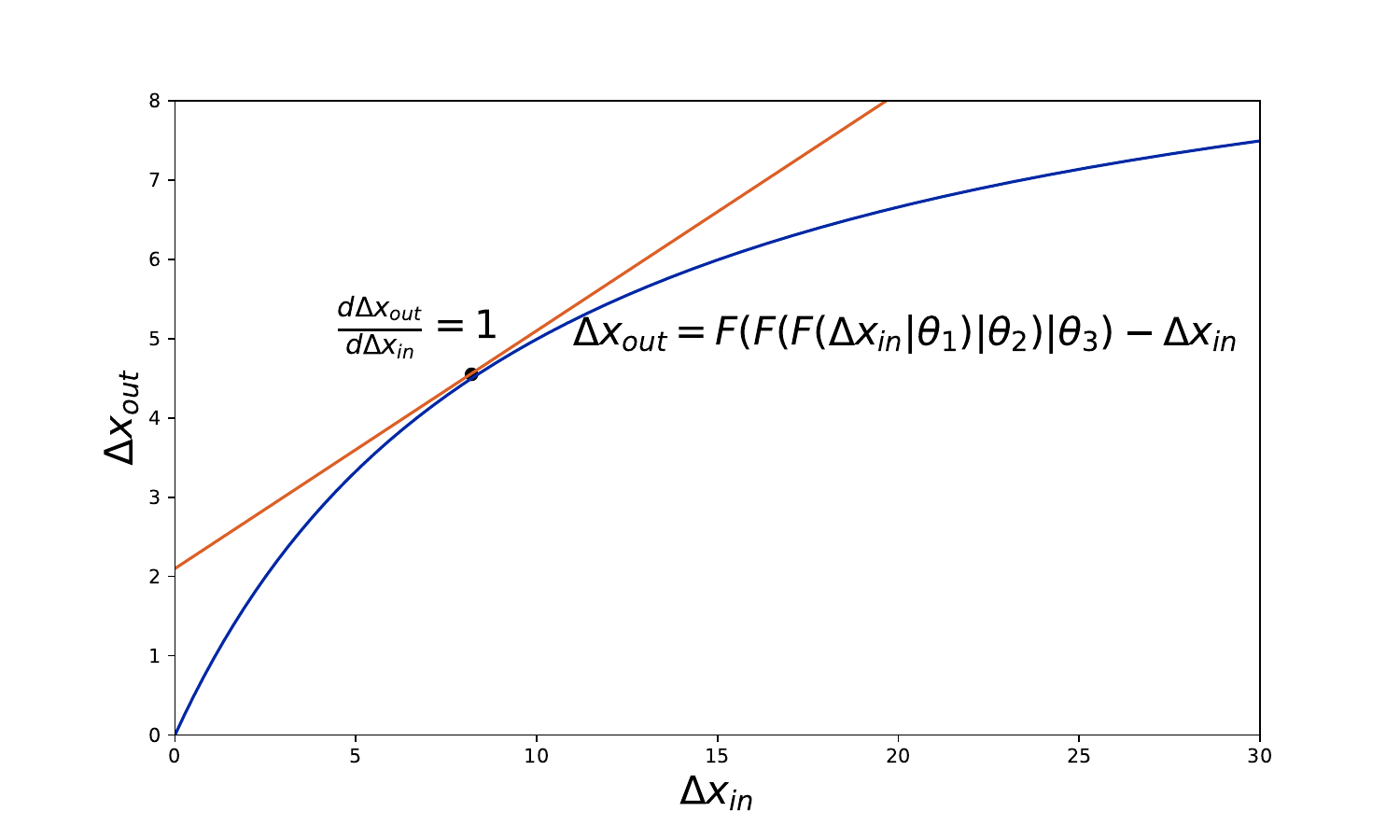}
    \caption{The arbitrage profit is maximal when $\frac{d\Delta x_{out}}{d\Delta x_{in}}=1$.}
    \label{fig:enter-label}
\end{figure}

Because of the convexity and monotonically-increasing property of this function, it is easy to use the bisection method to calculate the optimal input of $\Delta x_{in}$ or get the solution by solving the equation $\frac{d\Delta x_{out}}{d\Delta x_{in}}=1$.

We assume that $\Delta x_{in}^*$ is the optimal amount of input, then we have $\frac{d\Delta x_{out}}{d\Delta x_{in}}|_{\Delta x_{in}=\Delta x_{in}^*}=1$ and $log(p_{ij}^*) + log(p_{jk}^*) + log(p_{ki}^*)=0$, and there is no arbitrage anymore starting from token $X$. Because of $log(p_{ij}^*) + log(p_{jk}^*) + log(p_{ki}^*)=0$, there are also no arbitrage chances starting from other tokens in the loop, which means that $\frac{d\Delta y_{out}^,}{d\Delta y_{in}}|_{\Delta y_{in}=0}=1$ and $\frac{d\Delta z_{out}^,}{d\Delta z_{in}}|_{\Delta z_{in}=0}=1$ with new parameters $\theta_1^,$, $\theta_2^,$ and $\theta_3^,$.

By considering tokens' prices from CEXs, equation (\ref{e4}) has the same solution as equation (\ref{tra_obj}).
\begin{equation}
    \begin{split}
        Max[P_x(\Delta x_{out} - \Delta x_{in})]
    \end{split}
    \label{tra_obj}
\end{equation}
where $P_x$, $P_y$ and $P_z$ are tokens $X$, $Y$ and $Z$'s price from CEXs, respectively. 

The $MaxMax$ strategy to get the maximal monetized arbitrage profit by taking each token as input in turn can be expressed as
\begin{equation}
\begin{aligned}
    Max[Max[P_x(\Delta x_{out} - \Delta x_{in})],\\ Max[P_y(\Delta y_{out} - \Delta y_{in})], 
    Max[P_z(\Delta z_{out} - \Delta z_{in})]].\label{maxmax}
\end{aligned}
\end{equation}

\section{Mapping $MaxMax$ Strategy to $Convex \: Optimization$ Strategy}\label{optimization}

Now, the monetized arbitrage profit is measured in the same unit of fiat currency, which facilitates accumulating the monetized arbitrage profit from different tokens. By summing up the monetized arbitrage profits from different tokens, the object function in equation (\ref{tra_obj}) can be converted to a convex optimization problem with constraints $\Delta y_{out}^1 = \Delta y_{in}^2$ and $\Delta z_{out}^2 = \Delta z_{in}^3$  which is as follows: 
\begin{equation}
\begin{aligned}
\max \quad & P_x(\Delta x_{out}^3 - \Delta x_{in}^1)
+P_y(\Delta y_{out}^1 - \Delta y_{in}^2)+P_z(\Delta z_{out}^2 - \Delta z_{in}^3)\\
\textrm{s.t.} \quad 
&(x_1+\gamma \Delta x_{in}^1)\cdot(y_1 - \Delta y_{out}^1) \geq x_1\cdot y_1\\
&
(y_2+\gamma \Delta y_{in}^2)\cdot(z_2- \Delta z_{out}^2) \geq y_2\cdot z_2\\
&
(z_3+\gamma \Delta z_{in}^3)\cdot(x_3- \Delta x_{out}^3) \geq z_3\cdot x_3\\
&\Delta x_{out}^3 \geq \Delta x_{in}^1\\
&\Delta y_{out}^1 = \Delta y_{in}^2\\
&\Delta z_{out}^2 = \Delta z_{in}^3\\
&\Delta x, \Delta y,\Delta z \geq 0,
\end{aligned}
\label{optimization}
\end{equation}
where $\gamma=1-\lambda$, $x_1$, $y_1$, $x_2$, $y_2$, $y_3$, and $z_3$ are liquidity reserves in each pool and are constant parameters of the above equation.

Equation (\ref{optimization}) is a convex optimization problem with only one optimal solution. Conditions $\Delta y_{out}^1 = \Delta y_{in}^2$ and $
\Delta z_{out}^2 = \Delta z_{in}^3$ make it reduce to an initial problem, namely, adjusting the $\Delta x_{in}$ to maximize the profit $\Delta x_{out}-\Delta x_{in}$. Equation (\ref{optimization}) also implicitly indicates the transaction order of token $X\rightarrow Y \rightarrow Z \rightarrow X$.

If we change conditions $\Delta y_{out}^1 = \Delta y_{in}^2$ and $
\Delta z_{out}^2 = \Delta z_{in}^3$ to $\Delta y_{out}^1 \geq \Delta y_{in}^2$ and $\Delta z_{out}^2 \geq \Delta z_{in}^3$, the problem will become:
\begin{equation}
\begin{aligned}
\max \quad & P_x(\Delta x_{out}^3 - \Delta x_{in}^1)
+P_y(\Delta y_{out}^1 - \Delta y_{in}^2)+P_z(\Delta z_{out}^2 - \Delta z_{in}^3)\\
\textrm{s.t.} \quad 
&(x_1+\gamma \Delta x_{in}^1)\cdot(y_1 - \Delta y_{out}^1) \geq x_1\cdot y_1\\
&
(y_2+\gamma \Delta y_{in}^2)\cdot(z_2- \Delta z_{out}^2) \geq y_2\cdot z_2\\
&
(z_3+\gamma \Delta z_{in}^3)\cdot(x_3- \Delta x_{out}^3) \geq z_3\cdot x_3\\
&\Delta x_{out}^3 \geq \Delta x_{in}^1\\
&\Delta y_{out}^1 \geq \Delta y_{in}^2\\
&\Delta z_{out}^2 \geq \Delta z_{in}^3\\
&\Delta x, \Delta y,\Delta z \geq 0,
\end{aligned}
\label{optimization_up0}
\end{equation}
The problem in equation (\ref{optimization_up0}) is still a convex optimization problem because all constraint functions are still convex. However, the search space in the problem of equation (\ref{optimization_up0}) is larger than that in the problem of equation  (\ref{optimization}), so, we can get more profit in the problem of equation (\ref{optimization_up0}). In the search space of the problem of equation (\ref{optimization_up0}), we have constraints $\Delta x_{out}^3 \geq \Delta x_{in}^1$,
$\Delta y_{out}^1 \geq \Delta y_{in}^2$ and
$\Delta z_{out}^2 \geq \Delta z_{in}^3$, so, this arbitrage is also risky-free.

Now, if we delete the constraint conditions $\Delta x_{out}^3 \geq \Delta x_{in}^1$,
$\Delta y_{out}^1 \geq \Delta y_{in}^2$ and
$\Delta z_{out}^2 \geq \Delta z_{in}^3$ in equation (\ref{optimization_up0}), 



the search space in the new problem will be larger than that in the problem of equation (\ref{optimization_up0}), but both problems have the same object function, so, the maximal monetized arbitrage profit in the new problem will be larger or equal to that in the problem of equation (\ref{optimization_up0}). 
Removing constraint conditions $\Delta x_{out}^3 \geq \Delta x_{in}^1$, $\Delta y_{out}^1 \geq \Delta y_{in}^2$ and $\Delta z_{out}^2 \geq \Delta z_{in}^3$ further may lead to borrowing or shorting of some tokens, which may be very risky to investors, and we will not consider this corresponding convex optimization problem but only consider the problem of equation (\ref{optimization_up0}) in our research. 

The same logic also applies to $Max[P_y(\Delta y_{out} - \Delta y_{in})]$ and $Max[P_z(\Delta z_{out} - \Delta z_{in})]$, so, we can prove that the maximal monetized arbitrage profit $MaxMax$ strategy in equation (\ref{maxmax}) is equal to or less than that from the $Convex \: Optimization$ strategy in equation (\ref{optimization_up0}).

In this section, we take the loop with three tokens as an example, but the strategy of $MaxMax$ and $Convex \: Optimization$ can be applied to the loops with any length. In certain cases, arbitrageurs may only care more about how many specific tokens they can get. For example, if the objective function is $Max ({\Delta x_{out}^3 - \Delta x_{in}^1})$, then the convexity of the AMM constraint conditions will ensure that the conditions $\Delta y_{out}^1 = \Delta y_{in}^2$ and $\Delta z_{out}^2 = \Delta z_{in}^3$ in equation (\ref{optimization_up0}) are satisfied, then the problem in this case will reduce to equation (\ref{e4}).

Another significant question is whether we can get any risk-free arbitrage profit if no risk-free arbitrage profit can be found by traditional strategy or the $MaxMax$ strategy, namely when $\frac{d\Delta x_{out}}{d\Delta x_{in}}|_{\Delta x_{in}=0}=1$ or $log(p_{ij}^*) + log(p_{jk}^*) + log(p_{ki}^*)=0$, where the definitions of these variables are the same as that in Section \ref{traditional}. Now, we reason the answer to this question, here. Because of $\frac{d\Delta x_{out}}{d\Delta x_{in}}|_{\Delta x_{in}=0}=1$, we can get one unit of token $X$ if we input one unit of token $X$ into the arbitrage loop theoretically. Keeping any number of the other two tokens ($Y$ or $Z$) in hands as profit will lead to getting less number of token $X$, namely $\Delta x_{out} < \Delta x_{in}$, which is not consistent with the constraint condition $\Delta x_{out}^3 \geq \Delta x_{in}^1$ in equation \ref{optimization_up0}. So, we will have to keep $\Delta y_{out}^1 = \Delta y_{in}^2$ and $\Delta z_{out}^2 = \Delta z_{in}^3$ to ensure the condition $\Delta x_{out} \geq \Delta x_{in}$ is satisfied. Combining the condition $\frac{d\Delta x_{out}}{d\Delta x_{in}}|_{\Delta x_{in}=0}=1$ or $log(p_{ij}^*) + log(p_{jk}^*) + log(p_{ki}^*)=0$, it means that no more arbitrage profit can be gotten even if the $Convex \: Optimization$ strategy is applied. Overall, if the traditional strategies don't detect any arbitrage chances in a token loop, then the arbitrage optimization method can not either, namely the optimal solution by the convex optimization method is $(\Delta x_{out}^3, \Delta x_{in}^1,\Delta y_{out}^1, \Delta y_{in}^2,\Delta z_{out}^2, \Delta z_{in}^3)=(0,0,0,0,0,0)$.

\section{An Example of Compare the Profitability of $MaxMax$ Strategy, $MaxPrice$ Strategy and $Convex \: Optimization$ Strategy}\label{example}

This section gives an example of comparing the profitability of different strategies. Assuming the arbitrage loop $X\rightarrow Y \rightarrow Z \rightarrow X$ is profitable and $X$, $Y$ and $Z$ denote three different tokens. The liquidity pools are $(x,y) = (100,200)$, $(y,z) = (300,200)$, $(z,x) = (200,400)$, respectively. $x$, $y$, $z$ denote the reserves of tokens $X$, $Y$, and $Z$. The letter tuple denotes the reserve variables of tokens in the liquidity pool and the number tuple denotes the corresponding reserve of corresponding tokens in the pool.

Based on the token reserves in the above example, we can know that $\frac{PD_x}{PD_y}|_{(x,y)}=2$, $\frac{PD_y}{PD_z}|_{(y,z)}=\frac{2}{3}$, $\frac{PD_z}{PD_x}|_{(z,x)}=2$. $PD$ means the token's price from DEXs which is determined by the tokens' reserves in liquidity pools. The letter tuple $(x,y)$ in each formula denotes the liquidity pool that includes both tokens $X$ and $Y$, and other token letter tuples have a similar meaning. $\frac{PD_x}{PD_y}|_{(x,y)}=2$ denotes the ratio of the price of $X$ and $Y$ in liquidity pool $(x,y)$ is 2, and the other two formulas have the similar interpretations.

Because of $\frac{PD_x}{PD_y}|_{(x,y)}\cdot \frac{PD_y}{PD_z}|_{(y,z)}\cdot \frac{PD_z}{PD_x}|_{(z,x)}=\frac{8}{3}$ $(>1)$, we can get arbitrage profit in the loop $X\rightarrow Y \rightarrow Z \rightarrow X$.
Practitioners can also get arbitrage profit by inputting token $Y$ or token $Z$ and then getting more $Y$ or $Z$, respectively.

We assume the price of each token in a CEX is given, namely, $P_x=2\$$, $P_y=10.2\$$, $P_z=20\$$ and token $Z$ has the highest price when compared to token $X$ and $Y$. 


We assume traders only concentrate on the arbitrage starting from token $X$ and backing to token $X$ again ($X \rightarrow Y \rightarrow Z \rightarrow X$), but ignore the other two ways of arbitrage, such as $Y \rightarrow Z \rightarrow X \rightarrow Y$ and $Z \rightarrow X \rightarrow Y \rightarrow Z$, which may lead to getting less monetized arbitrage profit than the maximal monetized arbitrage profit they can get from this arbitrage loop in theory, which will be shown below.



\begin{figure}[!ht]
    \centering
    \includegraphics[width=0.35\textwidth]{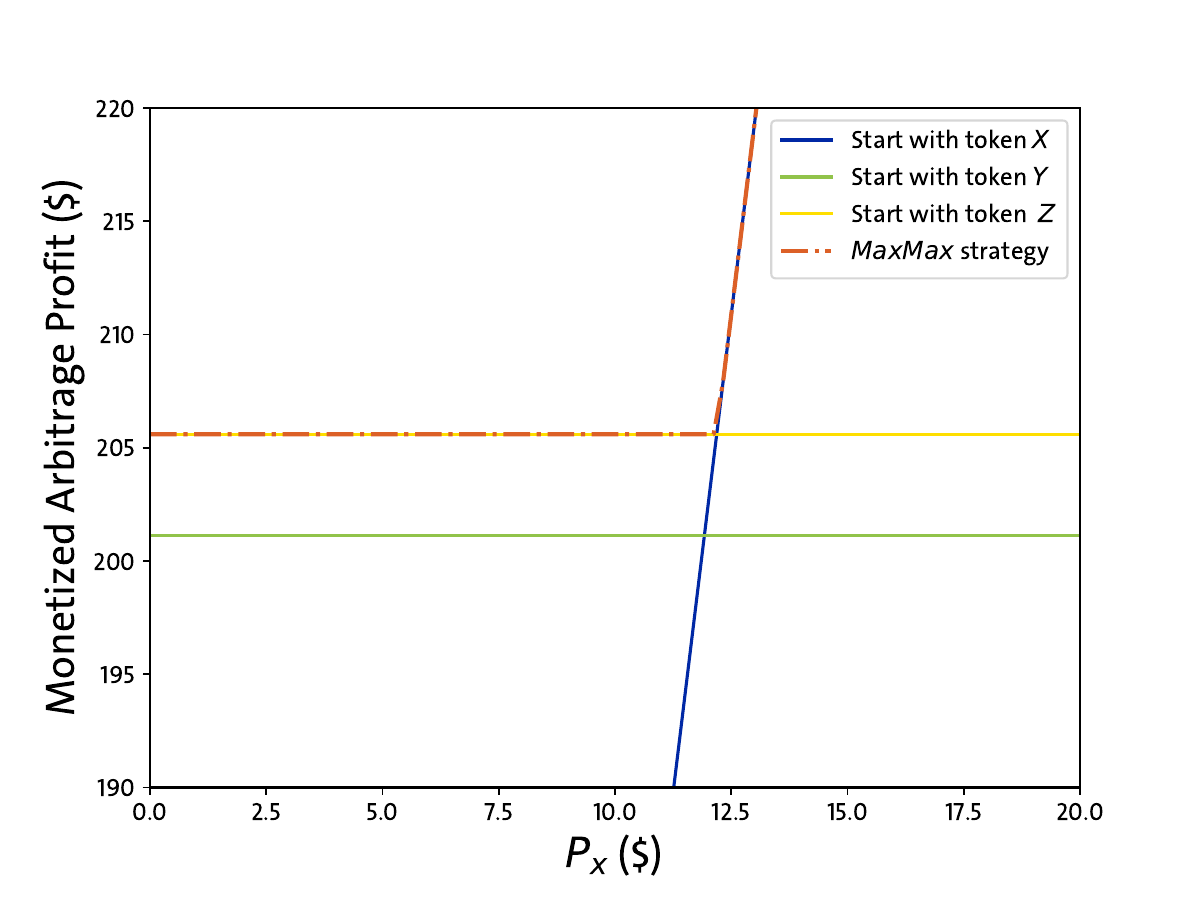}
    \caption{Monetized arbitrage profit by starting from different ways tokens and by the $MaxMax$ strategy with token $X$'s price from CEX ($P_x$) changing. The red dot line denotes the monetized arbitrage profit from the $MaxMax$ strategy, and the other three lines (green, yellow, and blue) denote the monetized arbitrage profits from the three different ways of arbitrage. 
    ``Start with token $X$" denotes the way of arbitrage $X \rightarrow Y \rightarrow Z \rightarrow X$ and the other two ``Start with $\cdots$" have similar interpretation.
    }
    \label{fig:maxmax}
\end{figure}

If we only consider the arbitrage path $X \rightarrow Y \rightarrow Z \rightarrow X$, the monetized arbitrage profit is only 33.7\$. However, the monetized arbitrage profit can be as high as 201.1\$ if we arbitrage by the order $Y \rightarrow Z \rightarrow X \rightarrow Y$ and 205.6\$ if by the order $Z \rightarrow X \rightarrow Y \rightarrow Z$, which makes it necessary to use the $MaxMax$ strategy in calculating the maximal monetized arbitrage profit. Fig. \ref{fig:maxmax} shows that the $MaxMax$ strategy is always the up-bound of the three ways of arbitrage when a token's price from CEX changes. Now, we analyze these different strategies in detail.

If the strategy starts from token $X$, then arbitrageurs can input 27.0 token $X$ and the arbitrage profit is 16.8 token $X$. If the strategy starts from token $Y$, then arbitrageurs can input 31.5 token $Y$ and the arbitrage profit is 19.7 token $Y$. If the strategy starts from token $Z$, then arbitrageurs can input 16.4 token $Z$ and the arbitrage profit is 10.3 token $Z$.

Fig. \ref{fig:maxmax} also shows that the $MaxPrice$ strategy (the yellow line) can not always get the most monetized arbitrage profit. For example, token $Z$'s price in CEXs is $P_z=20$\$, while when token $X$'s price in CEXs ($P_x$) is about 15\$, the monetized arbitrage profit from the way $X \rightarrow Y \rightarrow Z \rightarrow X$ is far above the $MaxPrice$ strategy $Z \rightarrow X \rightarrow Y \rightarrow Z$.

If we applied the $Convex \: Optimization$ for arbitrage, the monetized arbitrage profit can be up to 206.1\$ which is higher than that from the $MaxMax$ strategy. The $Convex \: Optimization$ strategy is as follows: inputting 31.3 token $X$ to get 47.6 token $Y$ in $(x,y)$ liquidity pool at first; then inputting 42.6 token $Y$ to get 24.8 token $Z$ in $(y,z)$ liquidity pool; at last, we input 17.1 token $Z$ to get 31.3 token $X$ again. The profit includes 5 token $Y$ and 7.7 token $Z$. The strategy can be implemented in any order. For example, we can input 42.6 token $Y$ to get 24.8 token $Z$ in $(y,z)$ liquidity pool; then, we input 17.1 token $Z$ to get 31.3 token $X$ in $(z,x)$ liquidity pool; at last, we input 31.3 token $X$ to get 47.6 token $Y$ in $(x,y)$ liquidity pool. To avoid any other risk, it is better to implement these three exchanges in the same transaction by applying flash loan in reality.

We can also find that the $Convex \: Optimization$ strategy needs to input more tokens compared to the $MaxMax$ strategy and other different ways of arbitrage strategies.

\begin{figure}[!ht]
    \centering
    \includegraphics[width=0.35\textwidth]{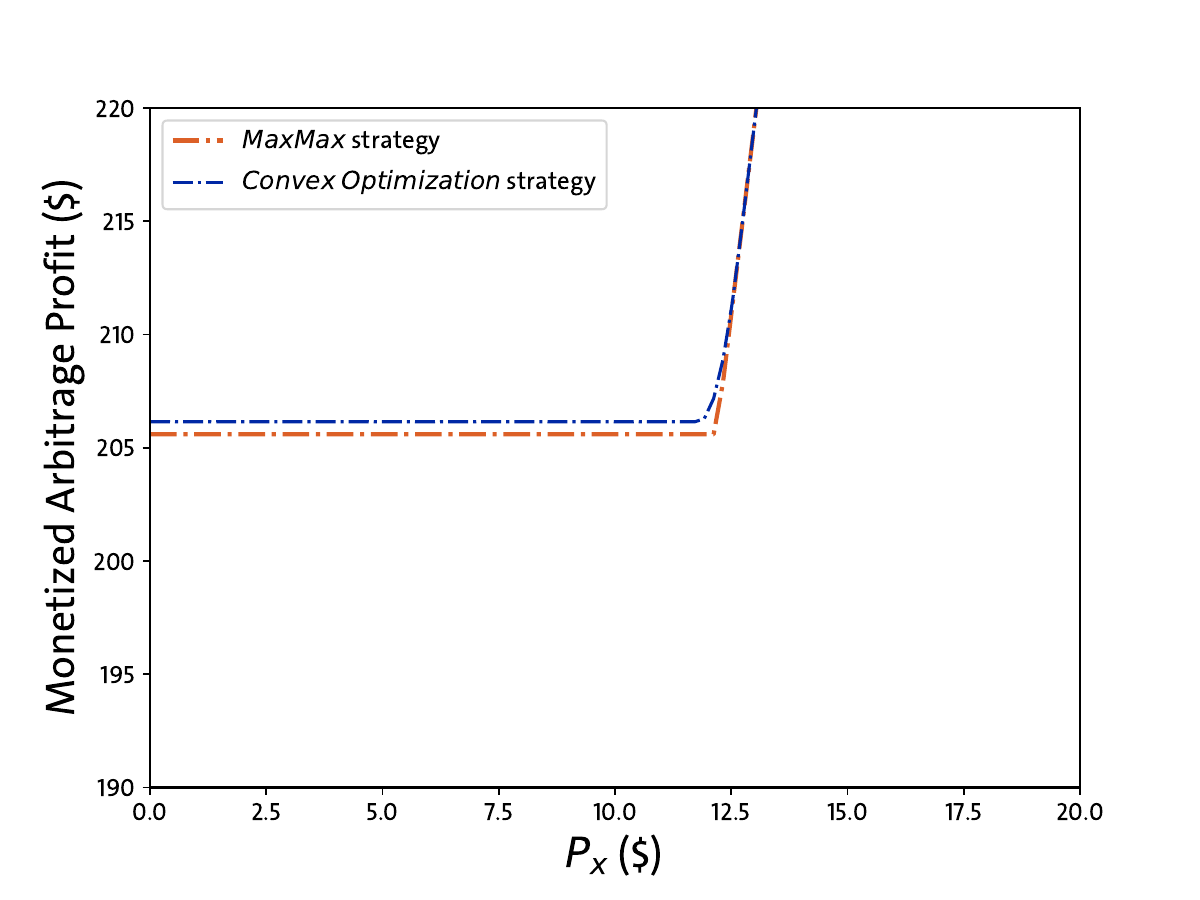}
    \caption{Monetized arbitrage profit by the $Convex \: Optimization$ strategy and the $MaxMax$ strategy with token $X$'s price from CEX ($P_x$) changing. The blue dot line denotes the monetized arbitrage profit by the $Convex \: Optimization$ strategy and the red dot line denotes the monetized arbitrage profit by the $MaxMax$ strategy. Arbitrageurs can always get more or at least equal profits by taking the $Convex \: Optimization$ strategy compared to the $MaxMax$ strategy.}
    \label{fig:opt}
\end{figure}

When the price of token $X$ changes from 0\$ to twenty dollars, it is easy to find that we can always get more or equal profit by $Convex \: Optimization$ strategy than that by the $MaxMax$ strategy in all cases, which is shown in Fig. \ref{fig:opt}.

\begin{figure}[!ht]
    \centering
    \includegraphics[width=0.4\textwidth]{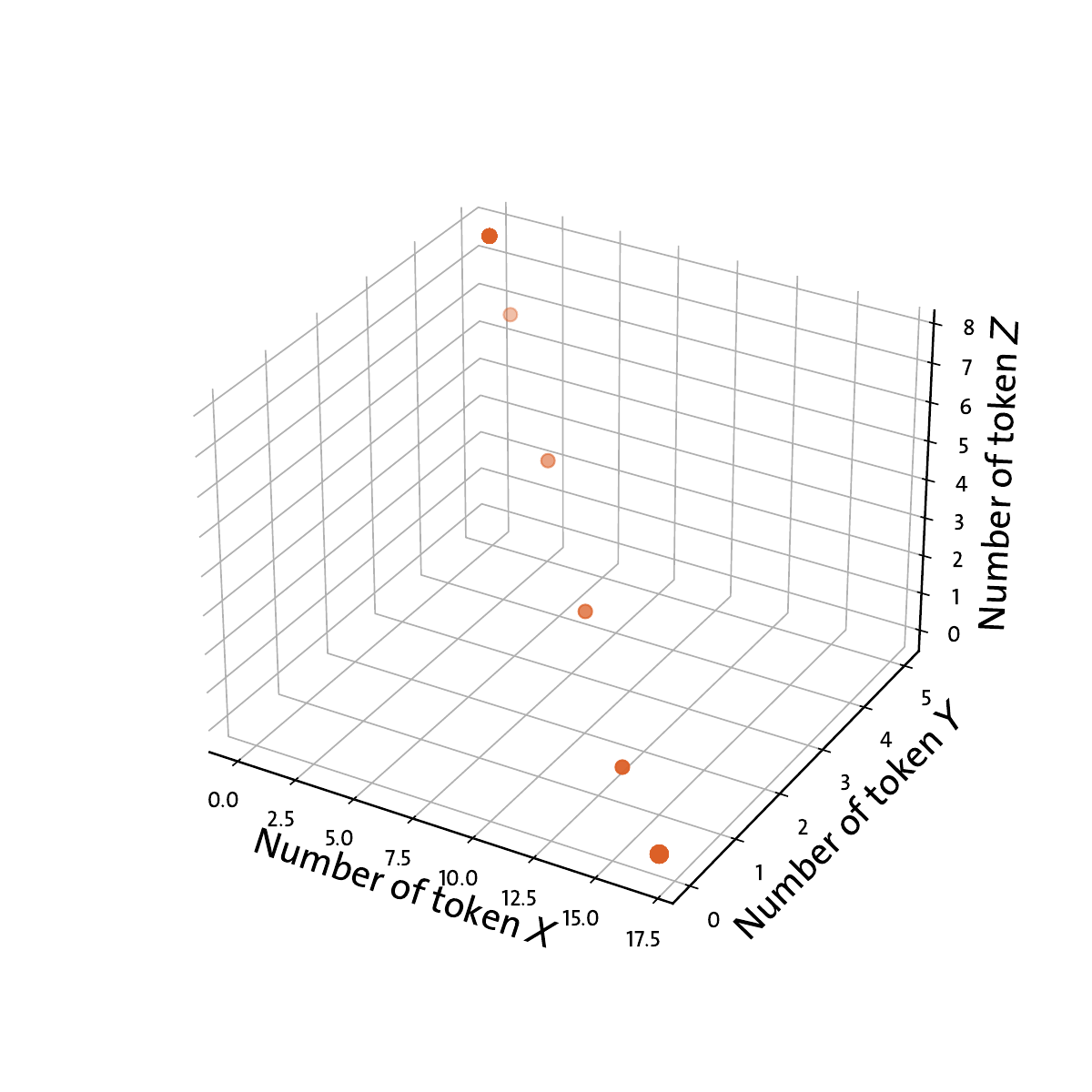}
    \caption{The profit in the form of the number of tokens $X$, $Y$, and $Z$ when token $X$'s price changes from 0 to twenty dollars with an interval of 0.2 dollars. The larger the point is in size, the more monetized arbitrage profit we get by the $Convex \: Optimization$ strategy.}
    \label{fig:opt_token}
\end{figure}

We also plot the arbitrage profit in the form of the net number of tokens $X$, $Y$, and $Z$ we get when token $X$'s price changes from 0 dollars to twenty dollars, as shown in Fig. \ref{fig:opt_token}. From Fig. \ref{fig:opt_token}, we can find that the optimal points mainly lie in six positions, which denotes that the optimization problem of our method is not linearly correlated to the price of tokens and is difficult to get a general analytical solution.

\section{Profitability Comparing Using Empirical Data on Uniswap V2}\label{uniswap_data}

We applied the liquidity pool data on September 1st, 2023, on Uniswap V2 and constructed the token exchange graph using similar procedures as that in the paper 'Identifying Arbitrage Paths and Loops on Decentralized Exchanges' Yu Zhang et al. (2024). The information on the token graph includes each token's reserve in corresponding liquidity pools. We chose those liquidity pools that have more than thirty thousand dollars TVL and where the number of each token is larger than one hundred. In the token graph we constructed, the node denotes the token, and the edge denotes the liquidity pool that includes the two different tokens corresponding to the edge's two ends. A liquidity pool (edge) is connected to another if they have the same token (node).

The token graph we got had fifty-one nodes (tokens) and two hundred and eight edges (liquidity pools).
The token's price from CEX (Binance) was downloaded on CoinGecko by API. To compare the profitability of different arbitrage strategies, we focus on only the loops with length 3 at first because the same analysis is easy to extend to arbitrage loops with any length.

We traversed all token loops with 3 tokens and selected those loops where arbitrage profit exists. For any loop, if the condition $p_{ij} \cdot p_{jk} \cdot p_{ki} >1$, or $log(p_{ij}) + log(p_{jk}) + log(p_{ki})>0$ is satisfied, then the loop is an arbitrage loop. In the above formula, $i$, $j$, $k$ are three different tokens, and $p_{ij}$ are the relative price between token $i$ and $j$ and is determined by their token reserve in the liquidity pool $(x,y)$. Other parameters in the above formula are defined similarly. After this procedure, we got one hundred and twenty-three arbitrage loops with length 3.

\begin{figure}[!ht]
    \centering
    \includegraphics[width=0.35\textwidth]{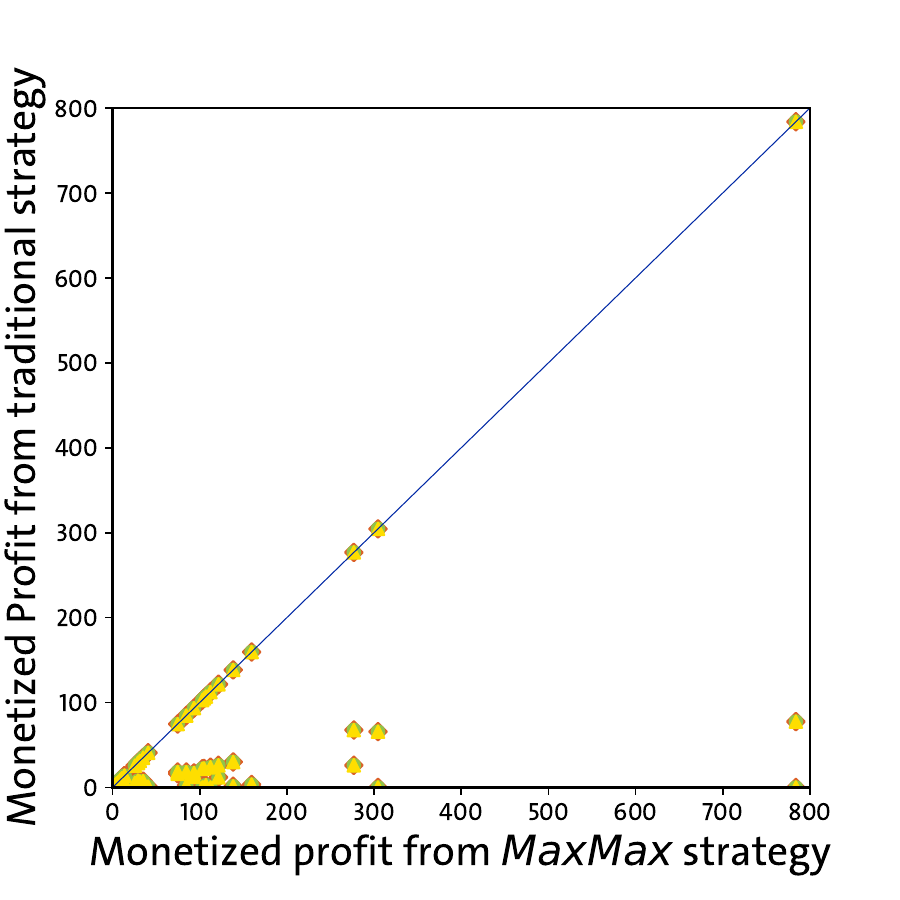}
    \caption{Monetized arbitrage profit from the $MaxMax$ strategy $vs$ that from traditional arbitrage strategies. The $x-$axis of each scatter plot is the monetized arbitrage profit from the $MaxMax$ strategy, and the $y-$axis of each scatter point denotes the monetized arbitrage profit from a traditional arbitrage strategy. An arbitrage loop with length 3 corresponds to three different points with different shapes (circle, diamond, triangle) and colors (yellow, red, green) but the same values in $x-$axis.}
    \label{fig:uniswap_profit_com}
\end{figure}

In this experiment, we record the respective monetized arbitrage profits from the traditional arbitrage strategies, the $MaxMax$ strategy, the $MaxPrice$ strategy, and the $Convex \: Optimization$ strategy. The monetized arbitrage profits from traditional strategies will correspond to several different values because the arbitrage may be implemented from any token in the arbitrage loop. For example, in a given arbitrage loop $X\rightarrow Y \rightarrow Z \rightarrow X$, the monetized arbitrage profits of traditional arbitrage strategies will include three different values which are the monetized arbitrage profit from the arbitrage order $X\rightarrow Y \rightarrow Z \rightarrow X$, $Y\rightarrow Z \rightarrow X \rightarrow Y$, and $Z\rightarrow X \rightarrow Y \rightarrow Z$, respectively. Then, we compare the arbitrage profit from the $MaxMax$ strategy and those from traditional strategies, which is shown in Fig. \ref{fig:uniswap_profit_com}.

All points in Fig. \ref{fig:uniswap_profit_com} are under or on the $45^\circ$ line, which means that the monetized arbitrage profits from the $MaxMax$ strategy are the up-bound of all other three traditional arbitrage strategies. This is easy to understand because the monetized arbitrage profits from the $MaxMax$ strategy equals the maximal monetized arbitrage profit from the traditional arbitrage strategies in definition. 

\begin{figure}[!ht]
    \centering
    \includegraphics[width=0.35\textwidth]{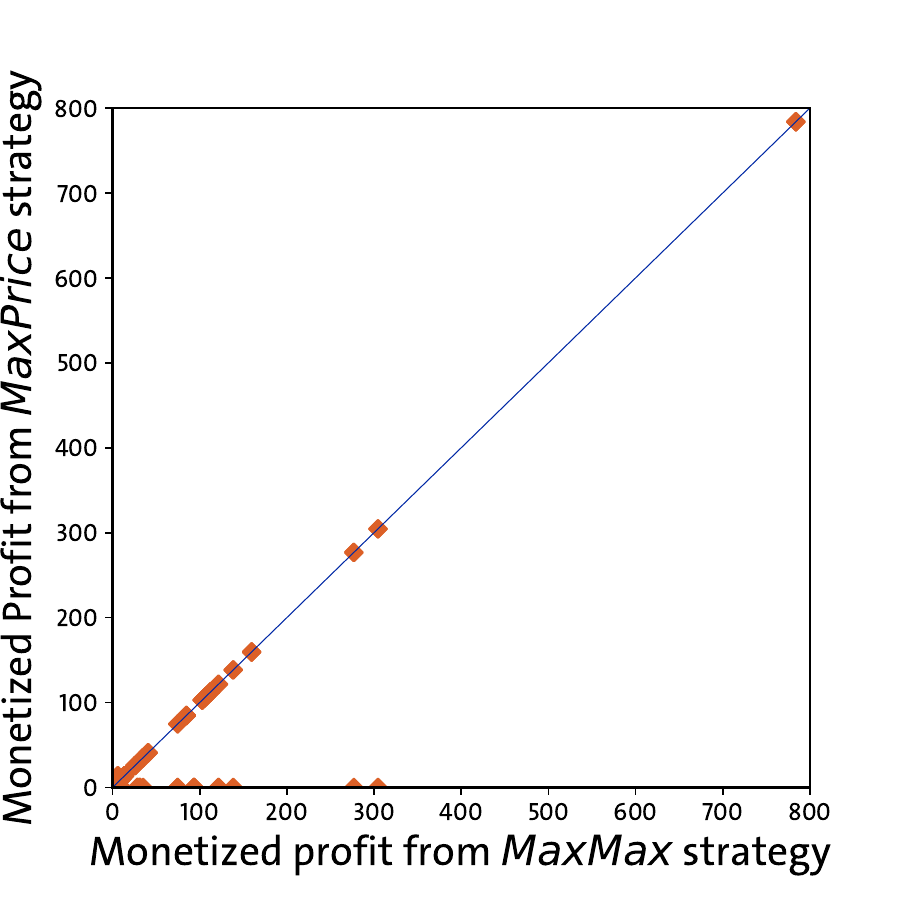}
    \caption{Monetized arbitrage profit from the $MaxPrice$ strategy $vs$ that from the $MaxMax$ strategy. The $x-$axis of each scatter plot is the monetized arbitrage profit from the $MaxMax$ strategy and the $y-$axis of each scatter point denotes the monetized arbitrage profit from the $MaxPrice$ strategy. One arbitrage loop corresponds to one red diamond point.}
    \label{fig:uniswap_profit_com_max_price}
\end{figure}

We compare the profitability of the $MaxMax$ strategy to the $MaxPrice$ strategy which is shown in Fig. \ref{fig:uniswap_profit_com_max_price}. From this figure, we can find that it is not always that the maximal monetized arbitrage profit can be obtained by starting arbitrage from the token with the highest price in CEXs, which makes the $MaxPrice$ strategy unreliable in calculating maximal monetized arbitrage profit.

\begin{figure}[!ht]
    \centering
    \includegraphics[width=0.35\textwidth]{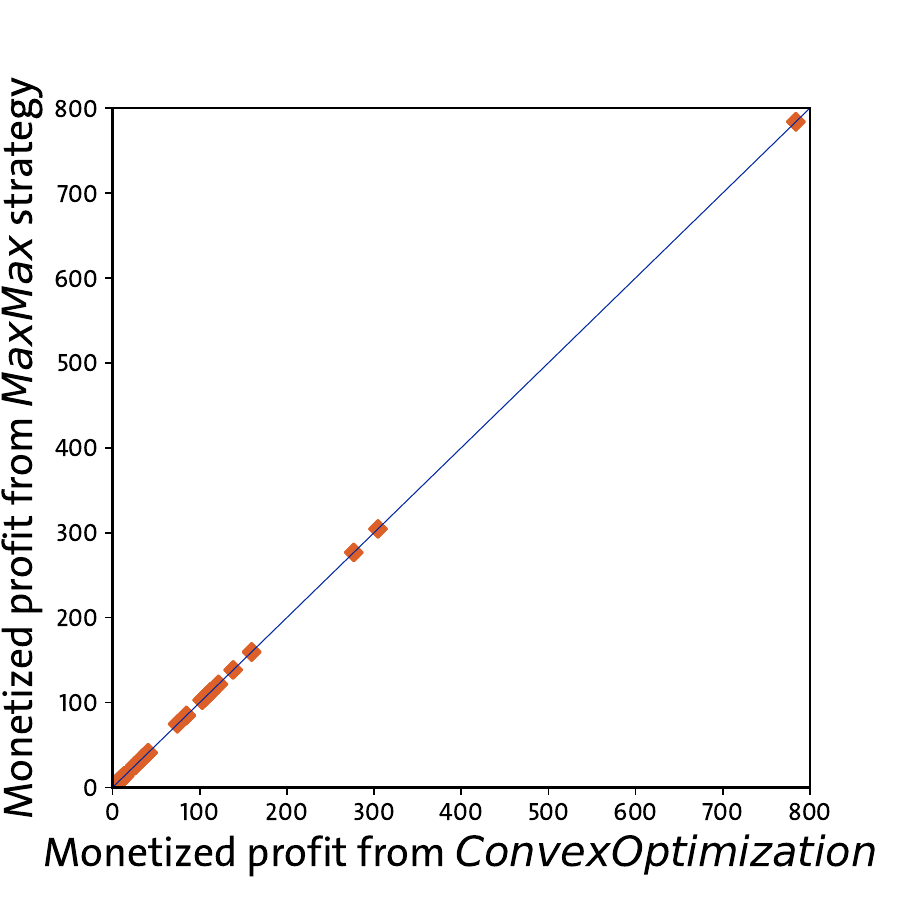}
    \caption{Monetized arbitrage profit from the $Convex \: Optimization$ strategy $vs$ that from the $MaxMax$ strategy. The $x-$axis of each scatter plot is the monetized arbitrage profit from the $Convex \: Optimization$ strategy and the $y-$axis of each scatter point denotes the monetized arbitrage profit from the $MaxMax$ strategy. One arbitrage loop corresponds to one red diamond point.}
    \label{fig:uniswap_profit_com_max}
\end{figure}

\begin{figure}[!ht]
    \centering
    \includegraphics[width=0.35\textwidth]{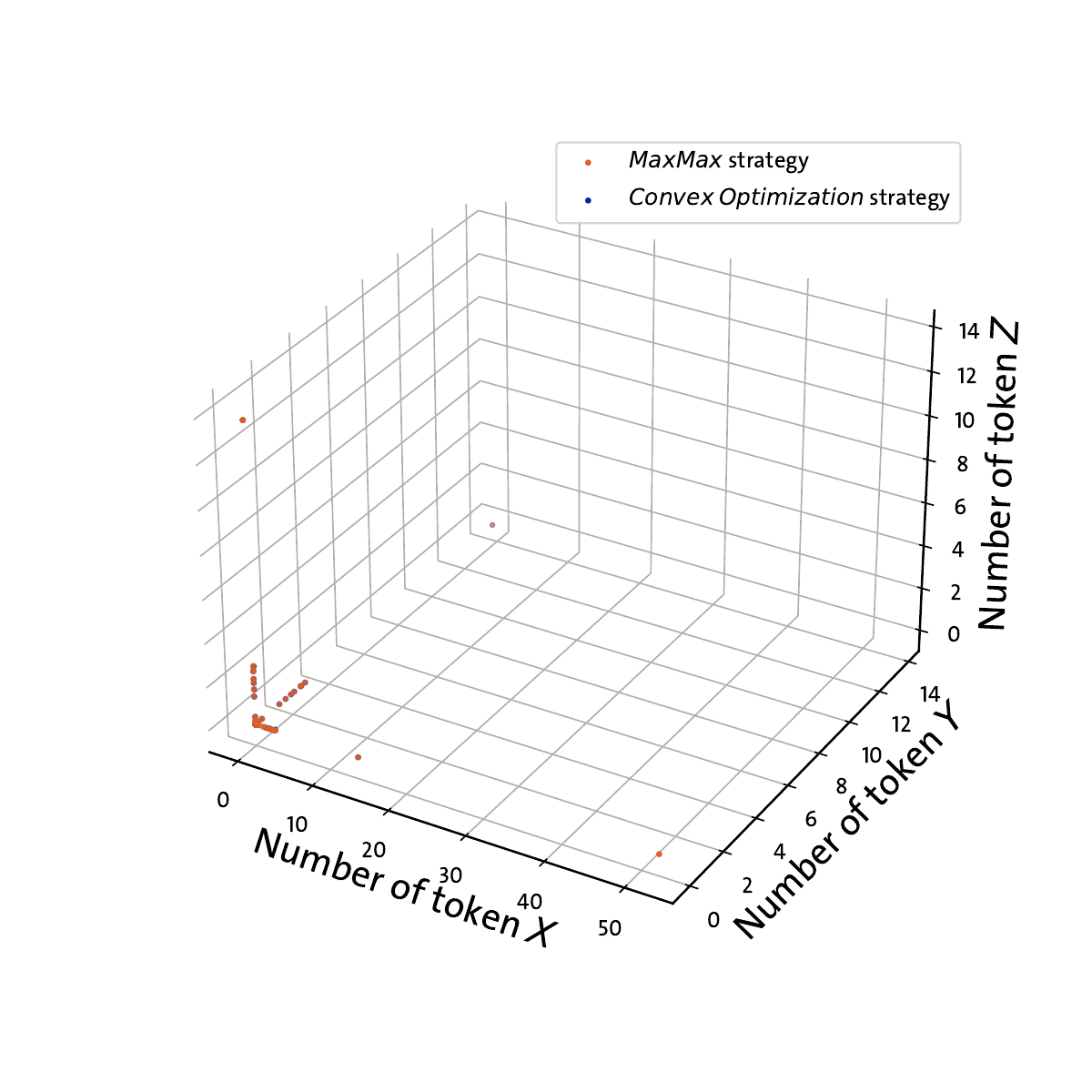}
    \caption{Arbitrage profit measured using the obtained net number of every token in the arbitrage loop, $Convex \: Optimization$ strategy $vs$ the $MaxMax$ strategy. In this figure, we plot the arbitrage profit in the unit of each token in the arbitrage loop. For an arbitrage loop with length 3, the arbitrage profit will include a specific number of each of these three tokens. We can find that the points from the $MaxMax$ strategy overlap with the points from the $Convex \: Optimization$ strategy.}
    \label{token_profit}
\end{figure}

The $MaxMax$ strategy can be mapped to the $Convex \: Optimization$ strategy and we have proved in theory that the $Convex \: Optimization$ strategy can get more monetized arbitrage profit than the $MaxMax$ strategy.
So, here, we also compare the profitability of the $Convex \: Optimization$ strategy and the $MaxMax$ strategy, by plotting the scatter plot which is shown in Fig. \ref{fig:uniswap_profit_com_max}. In Fig. \ref{fig:uniswap_profit_com_max}, all points are almost on the $45^\circ$ line, which means that the profit from the $Conevx \: Optimization$ strategy is almost equal to that from the $MaxMax$ strategy. This result may come partly from the reason that the monetized difference between the two arbitrage strategies is tiny. To find out whether these two strategies got the same results, we need to compare the arbitrage profits of these two arbitrage strategies from the perspective of the absolute net number of each token we get as profit. We find that the results are almost the same under the two strategies, which is shown in Fig. \ref{token_profit}.


\section{Result and Discussion}

In this paper, we put forward the $MaxMax$ strategy and the $Conevx \: Optimization$ strategy by considering tokens' prices from CEXs. It is easy to prove that the $MaxMax$ strategy can get more monetized arbitrage profit than the traditional arbitrage strategies and the $MaxPrice$ strategy. Then we mapped the $MaxMax$ strategy to the $Convex \: Optimization$ strategy and proved in theory that more monetized arbitrage profit can be obtained by the $Conevx \: Optimization$ strategy than the $MaxMax$ strategy. We also prove that if no arbitrage profit can be found by the traditional arbitrage strategies or the $MaxMax$ strategy, then no arbitrage profit can be detected by the $Convex \: Optimization$ strategy, either. These are some main theoretical contributions of this paper.


By the example and empirical data analysis, both the traditional strategies and the $MaxPrice$ strategy are not reliable if we focus on getting the maximal monetized arbitrage profit.
For the $MaxMax$ strategy and the $Convex \: Optimization$ strategy, both strategies get the maximal and almost equal monetized arbitrage profit in empirical data. For the $MaxMax$ strategy, we need to calculate the monetized arbitrage profit several times and then choose the maximal one. If an arbitrage loop has $n$ different tokens, then we need to calculate $n$ times. However, the calculation is simple and fast because the object function is convex and we can use the bisection method to get the optimal number of inputs and the arbitrage profit even if we need to calculate it multiple times. For example, for an arbitrage loop with a length of 10, the time required is in milliseconds level.
For the $Convex \: Optimization$ strategy, we only need to calculate once to get the monetized arbitrage profit, and the monetized arbitrage profit is at least equal to the monetized arbitrage profit from the $MaxMax$ strategy by the empirical analysis. However, the drawback of this $Convex \: Optimiazation$ is the high computation complexity for computers especially when the arbitrage loop is long. For example, for an arbitrage loop with a length of 10, the time required can be several seconds. While the average block time is 10 seconds, which makes the $Convex \: Optimization$ strategy not a perfect method in reality.

This paper proves that the monetized arbitrage profit from the $Convex \: Optimization$ strategy is equal to or larger than that from the $MaxMax$ strategy. However, we didn't give the discrepancy between these two kinds of strategies in theory, which can be a research direction in the future from the perspective of academic research.

\bibliographystyle{IEEEtran}
\bibliography{reference.bib}

\begin{thebibliography}{1}
\providecommand{\url}[1]{#1}
\csname url@samestyle\endcsname
\providecommand{\newblock}{\relax}
\providecommand{\bibinfo}[2]{#2}
\providecommand{\BIBentrySTDinterwordspacing}{\spaceskip=0pt\relax}
\providecommand{\BIBentryALTinterwordstretchfactor}{4}
\providecommand{\BIBentryALTinterwordspacing}{\spaceskip=\fontdimen2\font plus
\BIBentryALTinterwordstretchfactor\fontdimen3\font minus
  \fontdimen4\font\relax}
\providecommand{\BIBforeignlanguage}[2]{{%
\expandafter\ifx\csname l@#1\endcsname\relax
\typeout{** WARNING: IEEEtran.bst: No hyphenation pattern has been}%
\typeout{** loaded for the language `#1'. Using the pattern for}%
\typeout{** the default language instead.}%
\else
\language=\csname l@#1\endcsname
\fi
#2}}
\providecommand{\BIBdecl}{\relax}
\BIBdecl

\bibitem{Jani2022AnEmp}
\BIBentryALTinterwordspacing
J.~Berg, R.~Fritsch, L.~Heimbach, and R.~Wattenhofer, ``An empirical study of
  market inefficiencies in uniswap and sushiswap,'' \emph{Financial
  Cryptography Workshops}, 2022. [Online]. Available:
  \url{dblp.org/rec/journals/corr/abs-2203-07774}
\BIBentrySTDinterwordspacing

\bibitem{10.1145/3570639}
\BIBentryALTinterwordspacing
J.~Xu, K.~Paruch, S.~Cousaert, and Y.~Feng, ``Sok: Decentralized exchanges
  (dex) with automated market maker (amm) protocols,'' \emph{ACM Comput.
  Surv.}, vol.~55, no.~11, feb 2023. [Online]. Available:
  \url{https://doi.org/10.1145/3570639}
\BIBentrySTDinterwordspacing

\bibitem{cui2020detecting}
Z.~Cui, W.~Qian, S.~Taylor, and L.~Zhu, ``Detecting and identifying arbitrage
  in the spot foreign exchange market,'' \emph{Quantitative Finance}, vol.~20,
  no.~1, pp. 119--132, 2020.

\bibitem{yewa2023cyclic}
\BIBentryALTinterwordspacing
Y.~Wang, Y.~Chen, H.~Wu, L.~Zhou, S.~Deng, and R.~Wattenhofer, ``Cyclic
  arbitrage in decentralized exchanges,'' \emph{Arxiv:
  http://arxiv.org/abs/2105.02784v3}, 2023. [Online]. Available:
  \url{http://arxiv.org/pdf/2105.02784v3}
\BIBentrySTDinterwordspacing

\bibitem{zhou2021just}
L.~Zhou, K.~Qin, A.~Cully, B.~Livshits, and A.~Gervais, ``On the just-in-time
  discovery of profit-generating transactions in defi protocols,'' in
  \emph{2021 IEEE Symposium on Security and Privacy (SP)}.\hskip 1em plus 0.5em
  minus 0.4em\relax IEEE, 2021, pp. 919--936.

\bibitem{wang2022cyclic}
Y.~Wang, Y.~Chen, H.~Wu, L.~Zhou, S.~Deng, and R.~Wattenhofer, ``Cyclic
  arbitrage in decentralized exchanges,'' in \emph{Companion Proceedings of the
  Web Conference 2022}, 2022, pp. 12--19.

\bibitem{mclaughlin2023large}
R.~McLaughlin, C.~Kruegel, and G.~Vigna, ``A large scale study of the ethereum
  arbitrage ecosystem,'' in \emph{32nd USENIX Security Symposium (USENIX
  Security 23)}, 2023, pp. 3295--3312.

\bibitem{danos2021global}
V.~Danos, H.~E. Khalloufi, and J.~Prat, ``Global order routing on exchange
  networks,'' in \emph{Financial Cryptography and Data Security. FC 2021
  International Workshops: CoDecFin, DeFi, VOTING, and WTSC, Virtual Event,
  March 5, 2021, Revised Selected Papers 25}.\hskip 1em plus 0.5em minus
  0.4em\relax Springer, 2021, pp. 207--226.

\bibitem{berg2022empirical}
J.~A. Berg, R.~Fritsch, L.~Heimbach, and R.~Wattenhofer, ``An empirical study
  of market inefficiencies in uniswap and sushiswap,'' \emph{arXiv preprint
  arXiv:2203.07774}, 2022.

\end{thebibliography}

\appendix

\section{Experiment on Arbitrage Loops with Length 4}
\begin{figure}[!ht]
    \centering
    \includegraphics[width=0.3\textwidth]{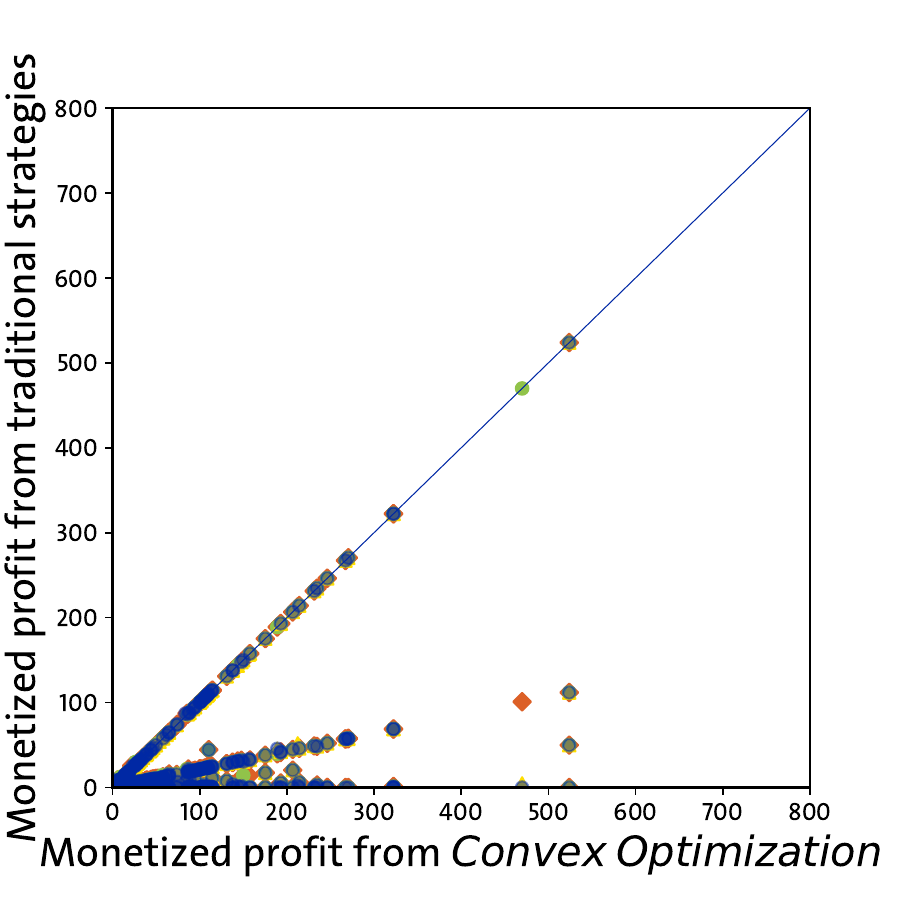}
    \caption{Monetized arbitrage profit, the $Convex \: Optimization$ strategy $vs$ the traditional strategies. The $x-$axis of each scatter plot is the monetized arbitrage profit from the $Convex \: Optimization$ strategy and the $y-$axis of each scatter point denotes the monetized arbitrage profit from other traditional arbitrage strategies. One arbitrage loop corresponds to four different points with different shapes but the same value in $x-$axis.}
    \label{fig:uniswap_profit_com4}
\end{figure}

\begin{figure}[!ht]
    \centering
    \includegraphics[width=0.3\textwidth]{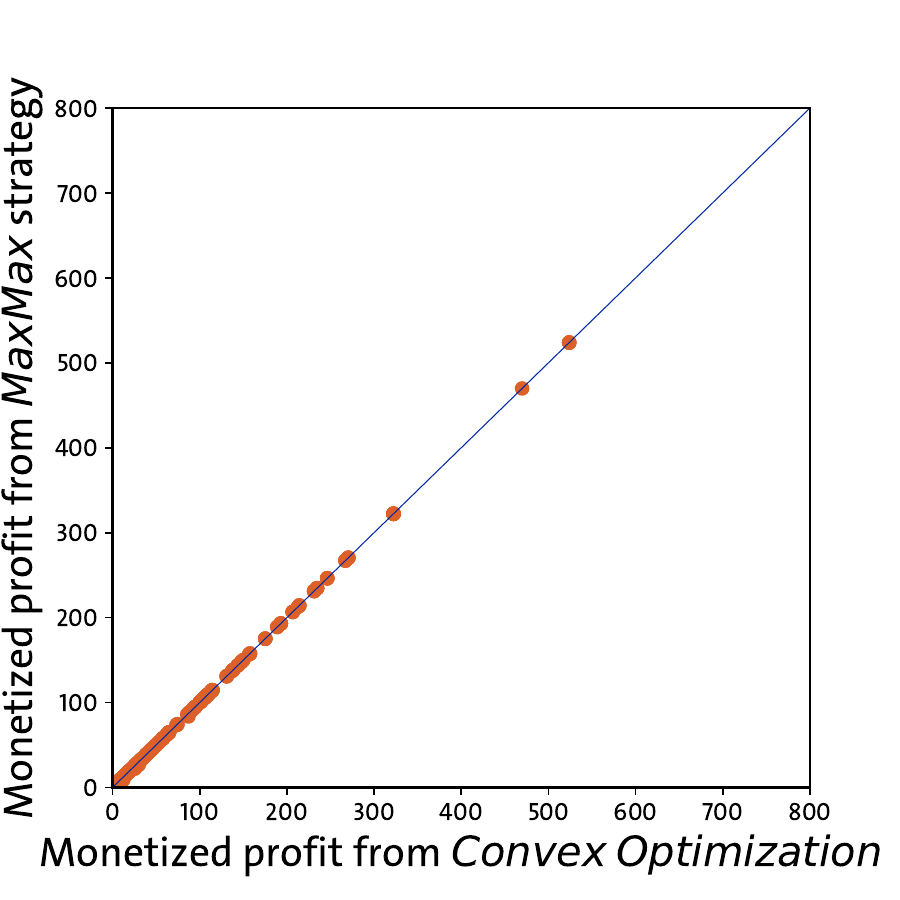}
    \caption{Monetized arbitrage Profit, the $Convex \: Optimization$ strategy $vs$ the $MaxMax$ strategy. The $x-$axis of each scatter plot is the monetized arbitrage profit from the $Convex \: Optimization$ strategy and the $y-$axis of each scatter point denotes the monetized arbitrage profit from the $MaxMax$ strategy. One arbitrage loop corresponds to one point in this figure.}
    \label{fig:uniswap_profit_com_max4}
\end{figure}

We repeat the same profit comparing procedure as that in Section \ref{uniswap_data} but focus on the arbitrage loop with length 4. The comparing results are similar to that in Section \ref{uniswap_data} and are shown in Fig. \ref{fig:uniswap_profit_com4} and \ref{fig:uniswap_profit_com_max4}.


\end{document}